\documentclass[twocolumn]{aastex62}
\usepackage[section]{placeins}
\usepackage{graphicx}
\usepackage{hyperref}
\usepackage[caption=false]{subfig}

\usepackage{float}
\usepackage{enumitem}

\shorttitle{Optical Analysis of HD96670}
\shortauthors{Gomez \& Grindlay}

\begin{document}
 
\title{Optical Analysis and Modeling of HD96670, a new Black Hole X-ray Binary Candidate}

\correspondingauthor{Sebastian Gomez}
\email{sgomez@cfa.harvard.edu}

\author[0000-0001-6395-6702]{Sebastian Gomez}
\affil{Center for Astrophysics \textbar{} Harvard \& Smithsonian, 60 Garden Street, Cambridge, MA, 02138, USA}

\author[0000-0002-1323-5314]{Jonathan E. Grindlay}
\affil{Center for Astrophysics \textbar{} Harvard \& Smithsonian, 60 Garden Street, Cambridge, MA, 02138, USA}

\begin{abstract}

We report on optical observations and modeling of HD96670, a single-line spectroscopic binary in the Carina OB2 association. We collected 10 epochs of optical spectroscopy, and optical photometry on 17 non-consecutive nights on the source. We construct a radial velocity curve from the spectra, and update the orbital period of the binary to be $P = 5.28388 \pm 0.00046$ days. The spectra show oxygen and helium absorption, consistent with an O-type primary. We see no evidence for spectral lines from the secondary star in the binary. We model the optical light curve and radial velocity curve simultaneously using the Wilson-Devinney code and find a best fit mass of $M_1 = 22.7^{+5.2}_{-3.6} M_\odot$ for the primary, and $M_2 = 6.2^{+0.9}_{-0.7} M_\odot$ for the secondary. An object of this mass is consistent with either a B-type star, or a black hole. Given that we see no absorption lines from the secondary, in combination with an observed hard power-law X-ray spectrum with $\Gamma = 2.6$ detected past 10 keV, maybe produced by wind accretion onto the secondary, we conclude that the secondary is most likely a black hole. We see asymmetrical helium lines with a shape consistent with the presence of a third star. If the secondary is indeed a black hole, this system would add to the small sample of only four possible black hole high mass X-ray binaries in the galaxy.

\end{abstract}



\section{Introduction} \label{sec:intro}

High mass X-ray binaries (HMXBs) are systems composed of a neutron star or black hole accreting matter from a high mass star typically $\gtrsim 10 M_\odot$, where accretion usually occurs via a focused wind from the massive primary to the compact secondary \citep{Charles06}. These systems live for a few million years, which is relatively short compared to their low mass X-ray binary counterparts, which can live for billions of years. Their short lifetimes contribute to the scarcity of detected HMXBs, particularly the ones with a black hole (BH) accretor. To date, there are only up to four known BH-HMXBs in the galaxy: Cygnus X-1 \citep{2011ApJ...742...84O}, SS 433 \citep{2010ApJ...722..586S}, Cygnus X-3 \citep{2013MNRAS.429L.104Z}, and MWC 656 \citep{2015MNRAS.452.2773G}. Cygnus X-1 is the only system with an undisputed black hole secondary. The nature of the compact object in the other three systems has been debated to be either a neutron star or black hole (e.g. \citealt{Goranskij19}). Most recently, \cite{Rivinius20} found the HR 6819 system to be consistent with a triple system made up by an outer Be star, and an inner B3 III star and a $> 5.0$ M$_\odot$ black hole. That being said, the presence of a black hole in this system has been highly debated (e.g., \citealt{Bodensteiner20, Badry20, Mazeh20}). Finding more BH-HMXBs is important for the study of stellar evolution and supernovae, and more recently, population studies of future gravitational wave sources; since BH-HMXBs are the systems that will eventually become NS-BH or BH-BH binaries \citep{Heuvel19}.

HD96670 is a single line spectroscopic binary (SB1) residing in the Carina OB2 association. \cite{2014ApJS..211...10S} classified the system as an O8.5(n)fp variable: a binary with broadened absorption lines and a rotational velocity of $v \sin(i) \sim 200$ km/s, strong NIII 4634, 4640, and 4642 emission, He II 4686 in emission above the continuum and a peculiar variable spectrum. \cite{2001Obs...121....1S} studied this source and determined an orbital period of $P = 5.52963 \pm 0.00018$ days and a systemic velocity of $\gamma = -9.0 \pm 1.7$ km/s from radial velocity measurements. More recently, \cite{2014ApJS..215...15S} obtained high-resolution optical images of O stars to search for their visible companions and found evidence for a third star in addition to the main binary in HD96670. This third star was detected at a projected separation from the main binary of $\rho = 29.9 \pm 3.37$ mas, and a magnitude difference to the main binary of $\Delta H = 1.26 \pm 0.10$ mag \citep{2014ApJS..215...15S}. The presence of the third star is not surprising for an high mass binary, given that \cite{2016arXiv160605347M} concluded that $\sim 35\%$ of stars in this mass range should be in triple systems and that $\sim30\%$ of OB stars are expected to have a compact remnant companion.

HD96670 was observed in the XMM-\textit{Newton} slew survey in 2010 and hard X-rays were detected \citep{Saxton08}, which motivated subsequent NuSTAR \citep{Harrison13} observations in 2015 (ObsID: 3000105000[2,4,6]). The X-ray spectrum is well modeled by a hard X-ray power-law with $\Gamma = 2.6 \pm 0.2$, with a corresponding X-ray luminosity of $L_x = (6.8\pm0.4) \times 10^{31}$ erg s$^{-1}$ in the 10-40 keV band (Grindlay et al. in prep).

The fact that the spectra of HD96670 is well described by a hard X-ray power-law and detected above 10keV, in combination with the measured mass of the secondary of $M_2 = 6.2$ M$_\odot$, as well as the detection He II 4686 in emission, and the fact that HD96670 resides in a dense OB association are all supportive of the idea that HD96670 harbors a black hole accretor. Since the location of HMXBs in the Milky Way correlate with the location of active OB associations \citep{Bodaghee12}.

This paper is structured as follows. In \S\ref{sec:data} we present the optical photometry and spectroscopy collected for this work, as well as a summary of the X-ray data presented in Grindlay et al., in prep. In \S\ref{sec:data_analysis} we outline the orbital features of the system, such as the orbital period, radial velocities, and presence of a third star. In \S\ref{sec:modeling} we describe the model of the light curve and radial velocity curve. We discuss the results in \S\ref{sec:Discussion}, and finally give our conclusions in \S\ref{sec:conclusions}.

\section{Observations} \label{sec:data}

\subsection{Optical Photometry} \label{subsec:photometry}

\begin{figure}
	\centering
	\includegraphics[width=\columnwidth]{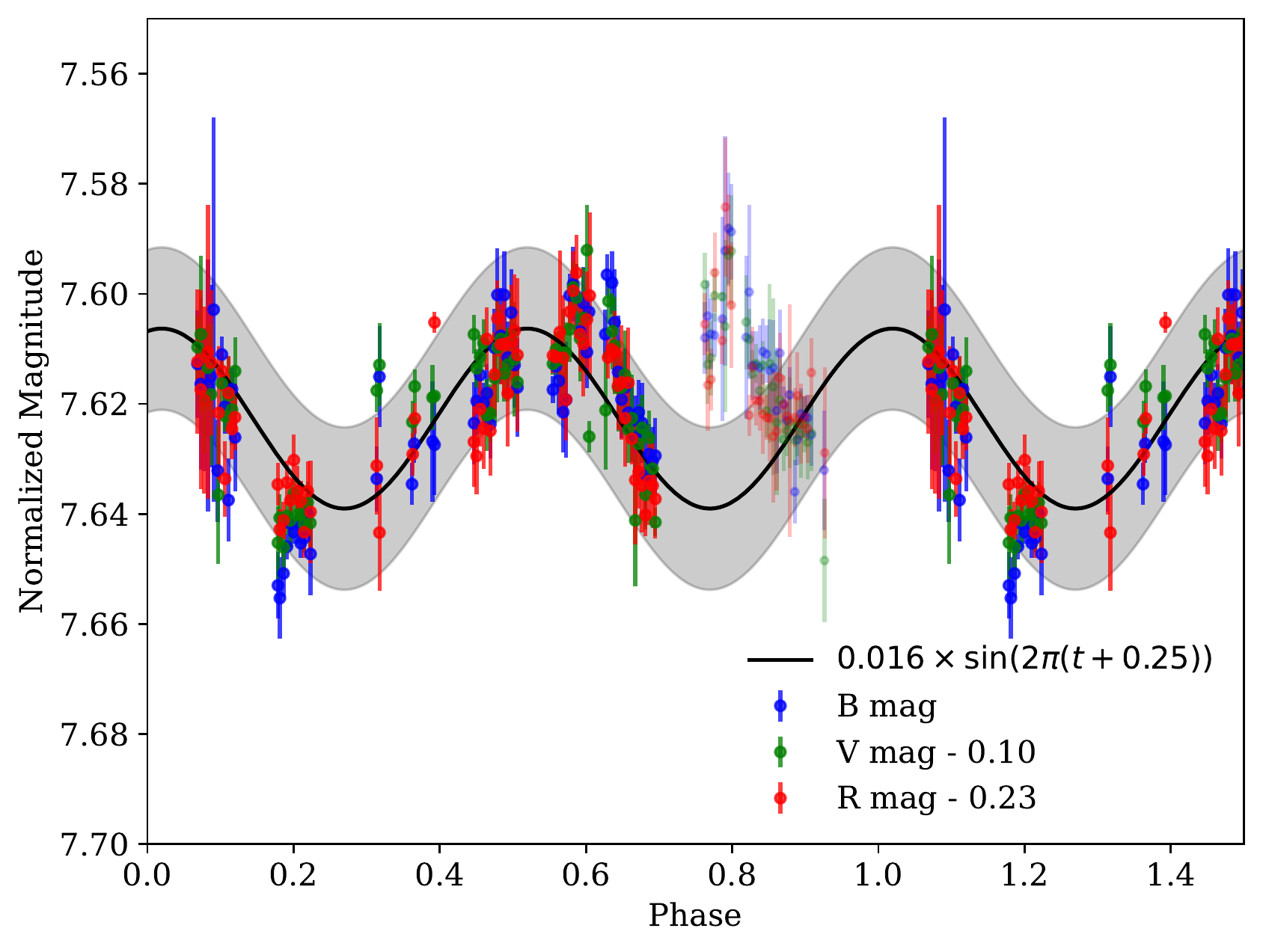}
	\caption{AAVSO and CTIO photometry of HD96670 phased at the orbital period of 5.28388 days. Each band has been normalized to a common average to make the sinusoidal shape apparent. The faint points in the phase range $0.75-0.95$ diverge from the light curve shape expected from ellipsoidal modulations. This excess could be caused by either a narrow focused wind onto the secondary, a hot spot, or a decrease in the optical thickness of the wind. \label{fig:photometry}}
\end{figure}

We requested optical photometry from the American Association of Variable Star Observers (AAVSO), and observed HD96670 in the B, V, and R bands between 2015 June and 2015 July with the Berry 6cm bright star monitor telescope, located in Perth, Australia. We obtained additional photometry with the SMARTS CTIO 1.3m Telescope using ANDICAM \citep{DePoy03} also in the B, V, and R bands between 2016 March and 2016 April. The AAVSO data were reduced and aperture photometry extracted with the AAVSO dedicated pipeline. The CTIO images were bias-subtracted and flat-fielded, and aperture photometry was extracted using standard routines with IRAF\footnote{\label{IRAF}IRAF is written and supported by the National Optical Astronomy Observatories, operated by the Association of Universities for Research in Astronomy, Inc. under cooperative agreement with the National Science Foundation.}. The resulting light curve is shown in Figure \ref{fig:photometry}, and a log of photometric observations is provided in Table~\ref{tab:photometry}.

\begin{deluxetable*}{ccccc}
    \tablecaption{Optical Photometry of HD96670 \label{tab:photometry}}
    \tablewidth{0pt}
    \tablehead{
    \colhead{UT Date}   & \colhead{B-band} & \colhead{V-band} & \colhead{R-band} & \colhead{Telescope} \\
                        & \colhead{(s)}    & \colhead{(s)}    & \colhead{(s)}    & }
    \startdata
    2015 Jun 9          & $104\times60$    & $104\times30$    & $104\times30$    & AAVSO Berry     \\
    2015 Jun 10         & $99 \times60$    & $99 \times30$    & $99 \times30$    & AAVSO Berry     \\
    2015 Jun 12         & $27 \times50$    & $27 \times30$    & $27 \times20$    & AAVSO Berry     \\
    2015 Jun 22         & $11 \times50$    & $11 \times30$    & $11 \times20$    & AAVSO Berry     \\
    2015 Jun 23         & $82 \times50$    & $82 \times30$    & $82 \times20$    & AAVSO Berry     \\
    2015 Jun 24         & $79 \times50$    & $79 \times30$    & $79 \times20$    & AAVSO Berry     \\
    2015 Jun 25         & $89 \times50$    & $89 \times30$    & $89 \times20$    & AAVSO Berry     \\
    2015 Jul 11         & $44 \times50$    & $44 \times30$    & $44 \times20$    & AAVSO Berry     \\
    2015 Jul 12         & $64 \times50$    & $64 \times30$    & $64 \times20$    & AAVSO Berry     \\
    2015 Jul 14         & $62 \times50$    & $62 \times30$    & $62 \times20$    & AAVSO Berry     \\
    2016 Mar 16         & $8  \times0.2$   & $8 \times0.2$    & $8  \times0.2$   & CTIO ANDICAM    \\
    2016 Mar 22         & $8  \times0.2$   & $9 \times0.2$    & $8  \times0.2$   & CTIO ANDICAM    \\
    2016 Mar 25         & $32 \times0.4$   & $29\times0.4$    & $29 \times0.4$   & CTIO ANDICAM    \\
    2016 Mar 31         & $8  \times0.8$   & $8 \times0.8$    & $8  \times0.8$   & CTIO ANDICAM    \\
    2016 Apr 2          & $8  \times0.8$   & $9 \times0.8$    & $9  \times0.8$   & CTIO ANDICAM    \\
    2016 Apr 3          & $16 \times0.8$   & $16\times0.8$    & $16 \times0.8$   & CTIO ANDICAM    \\
    2016 Apr 4          & $16 \times0.8$   & $16\times0.8$    & $16 \times0.8$   & CTIO ANDICAM    \\
    \enddata
    \tablecomments{Log of photometric observations performed wtih the CTIO 1.3m telescope and the AAVSO Berry telescope. With the individual exposure times for each filter shown in seconds.}
\end{deluxetable*}

\subsection{Optical Spectroscopy} \label{subsec:spectra}

We collected optical spectroscopy with the CHIRON echelle spectrograph on the CTIO 1.5m SMARTS telescope \citep{CHIRON, Tokovinin13}, the Inamori-Magellan Areal Camera and Spectrograph (IMACS; \citealt{dressler11}) spectrograph on the Magellan Baade Telescope at Las Campanas Observatory, and the MagE echellette spectrograph on the same telescope \citep{MagE}, a log of observations with the corresponding grating and wavelength range of each spectrograph is shown in Table \ref{tab:spectroscopy}. The CHIRON data were reduced using the dedicated CHIRON data reduction pipeline. The MagE data were reduced using the MagE data reduction pipeline described in \cite{Kelson03}. The IMACS data were reduced using standard IRAF$^{\ref{IRAF}}$ routines using the {\tt twodspec} and {\tt apall} packages. The data were bias-subtracted and flat-fielded, we then modeled and subtracted the sky emission, and finally optimally extract the spectra and calibrate the wavelength to the lines from an arc lamp taken directly after each spectrum.

\subsection{Distance} \label{sec:distance}

The parallax of HD96670 as measured by Gaia DR2 is $\pi = 0.248 \pm 0.034$ mas, which corresponds to a distance of $d = 3.87^{+0.64}_{-0.43}$ kpc \citep{2018arXiv180409376L}. Nevertheless, it is noted that Gaia parallax measurements for binary stars can be unreliable, since the orbital movement of the stars can be confused with an apparent parallax. Therefore, we instead gather a sample of 12 bright O stars similar to HD96670 in the Carina OB2 association and use their average distance (weighted by their uncertainty) and infer a distance to HD96670 of $d = 2.83^{+1.3}_{-0.80}$ kpc, which we adopt throughout this work.

\begin{deluxetable*}{cccccc}
    \tablecaption{Optical Spectroscopy of HD96670 \label{tab:spectroscopy}}
    \tablewidth{0pt}
    \tablehead{
    \colhead{UT Date}   & \colhead{Exposure} & \colhead{Telescope +} & \colhead{Grating} & \colhead{Resolution} & \colhead{Wavelength range} \\
                        & \colhead{(s)}      & \colhead{Instrument}  & (lines/mm)        & (\AA)                & (\AA)}
    \startdata
    2016 Apr 22         & $12 \times 600$    & CTIO + CHIRON         & R80000            & 0.11                 & 4580-8760                 \\
    2016 Apr 26         & $6  \times 900$    & CTIO + CHIRON         & R80000            & 0.11                 & 4580-8760                 \\
    2016 Apr 27         & $3  \times 1800$   & CTIO + CHIRON         & R80000            & 0.11                 & 4580-8760                 \\
    2016 May 21         & $6  \times 1000$   & CTIO + CHIRON         & R80000            & 0.11                 & 4580-8760                 \\
    2016 May 22         & $6  \times 1000$   & CTIO + CHIRON         & R80000            & 0.11                 & 4580-8760                 \\
    2016 May 26         & $6  \times 1000$   & CTIO + CHIRON         & R80000            & 0.11                 & 4580-8760                 \\
    2017 Jan 29         & $1  \times 43$     & Magellan + MagE       & R4100             & 0.6                  & 3100-8300                 \\
    2017 Jan 31         & $3  \times 30$     & Magellan + IMACS      & R1200             & 1.5                  & 3869-6753                 \\
    2017 Feb 1          & $4  \times 30$     & Magellan + IMACS      & R1200             & 1.5                  & 3869-6753                 \\
    2017 Feb 2          & $3  \times 30$     & Magellan + IMACS      & R1200             & 1.5                  & 3869-6753                 \\
    \enddata
	\tablecomments{Spectroscopic observations gathered with the CTIO 1.5m CHIRON spectrograph, MagE, and IMACS spectrogrphs on the Baade Telescope at the Magellan Observatory.}
\end{deluxetable*}

\subsection{X-ray Observations}

HD96670 was first detected in the X-rays by the XMM-\textit{Newton} slew survey on 2010 July 23 with a luminosity of \mbox{$(2.4 \pm 1.3)\times10^{34}$ erg s$^{-1}$} in the 0.2-12 keV band \citep{Saxton08}. Subsequent observations by NuSTAR on 2015 March 04, March 29, and April 29 revealed an X-ray spectrum with a hard power-law component.

Grindlay et al. in prep. fit the NuSTAR data with an apec spectrum (a model for collisionally-ionized diffuse gas) and find a model with a best fit temperature of $kT = 5.2^{+1.8}_{-0.9}$ keV and a corresponding luminosity of \mbox{$L_x = 1.7\times10^{32}$ erg s$^{-1}$}. This temperature is much too high for either an isolated O star or one in a colliding wind binary. For either case, typical temperatures are $< 2$keV \citep{2012ASPC..465..301G} and O8-O9V stars have soft X-ray luminosities $< 10^{31}$erg/s \citep{2011ApJS..194....7N}. Instead, the X-ray spectra is best fit with a hard X-ray power-law, a feature commonly observed in X-ray binaries. The best fit power-law index is $\Gamma = 2.6 \pm 0.2$, with a corresponding X-ray luminosity of $L_x = (2.2\pm0.2) \times 10^{32}$ erg s$^{-1}$ in the 2-10 keV band, and $L_x = (6.8\pm0.4) \times 10^{31}$ erg s$^{-1}$ in the 10-40 keV band (Grindlay et al. in prep). For these calculations we adopt a distance of $d = 2.83$ kpc.

The hard X-ray component is hard to explain with a non-interacting binary. Motivating the idea that the secondary in the binary is actually a black hole. The best fit power-law of $\Gamma = 2.6$. For reference, Cygnus X-1 in the hard state has a power law index of $\Gamma \sim 1.85$ \citep{Pottschmidt03}, while in the soft state it has a power law index $\Gamma \sim 2.5$ \citep{2014ApJ...790...29G}.

Additionally, we measure the luminosity of the primary assuming a temperature of 38,000K \citep{Hohle10} and a primary radius of $17.1$ R$_\odot$ (Derived in \S\ref{sec:modeling}) to be $(2.1 \pm 0.4)\times 10^{39}$ erg s$^{-1}$, where we adopted an extinction to HD96670 of $A_V = 1.497$, estimated by \cite{Apellaniz18}. If we instead fit a blackbody curve to the 2MASS \citep{Skrutskie06}, WISE \citep{Cutri12}, and optical \citep{HIPPARCOS97} photometry we obtain a more uncertain, but consistent, value for the luminosity of the primary of $(5.1 \pm 2.9)\times 10^{39}$ erg s$^{-1}$. The ratio of X-ray ($0.5 - 10$ keV range) to optical luminosity is therefore $\log(L_x / L_{\rm opt}) = -6.5 \pm 0.1$, typical for values between of O and B stars from \cite{Naze14}.

\section{Data Analysis} \label{sec:data_analysis}

\subsection{Third Star}\label{sec:3rdstar}

We know HD96670 is a single line spectroscopic binary from radial velocity measurements \citep{2001Obs...121....1S}. In addition to this, \cite{2014ApJS..215...15S} obtained high-resolution adaptive optics images of HD96670 when searching for companions to massive O-type stars and found evidence for a third star in proximity to the main binary at a projected separation of $\rho = 29.9 \pm 3.37$ mas and a magnitude difference in the $H$ band of $\Delta H = 1.26 \pm 0.10$ mag to that of the main binary.

\begin{figure}
	\centering
	\includegraphics[width=\columnwidth]{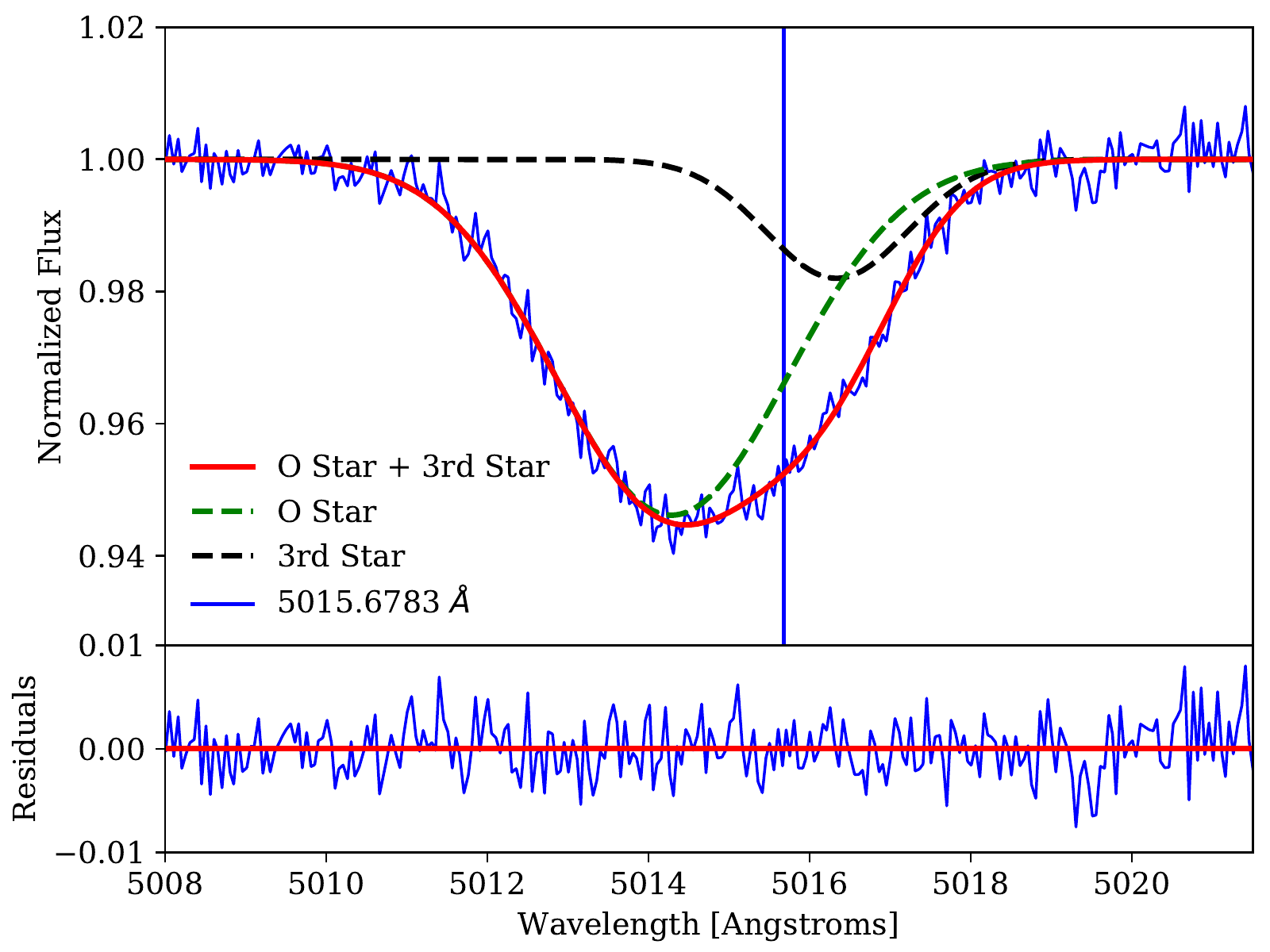}
	\caption{Absorption line of He 5015, clearly showing an asymmetrical shape. We fit this shape with a double-Gaussian model, one Gaussian (green) accounts for the contribution from the primary O star, and the second Gaussian (black) accounts for the contribution from the third star in the system. The red line is the sum of these two components and the blue vertical line shows the rest wavelength of He I 5015. \label{fig:HeI5015}}
\end{figure}

The optical spectra of HD96670 show strong helium and oxygen absorption lines from the O-star primary. These lines appear asymmetric, yet we see no evidence for absorption lines that would be consistent with the expected velocity of the secondary. Instead, the shape of the lines can be well modeled by a double-Gaussian model in which one Gaussian tracks the absorption line from the primary O-star, and the second Gaussian is at a fixed radial velocity, consistent with the third star interpretation. In Figure~\ref{fig:HeI5015} we show the best example of the asymmetric helium lines, where the contribution from the primary is shown in green, and the contribution from the third star in shown in black. We find a best fit of $v = 39.3 \pm 0.3$ km s$^{-1}$ for velocity of the third star. The flux ratio of the Gaussian component from the third star to the component from the O star primary is $D_{3rd}/D_{O} = 0.3344 \pm 0.0065$.

\begin{figure*}
	\includegraphics[width=\textwidth]{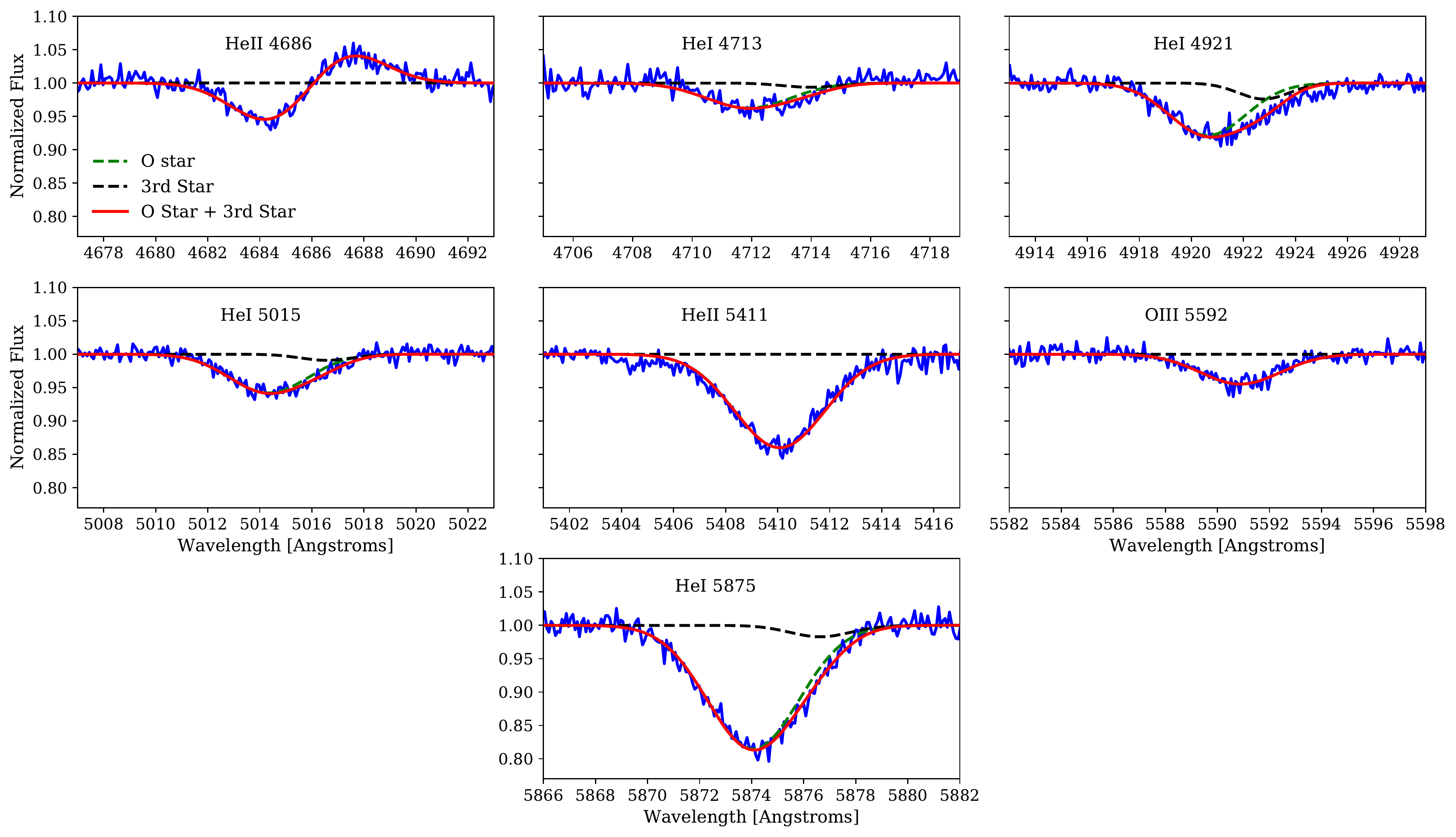}
	\caption{The seven strongest absorption lines of HD96670 from one spectrum taken at a pahse of 0.97. The He I lines were fit with a double Gaussian function, where the green Gaussian accounts for the contribution from the primary O star, and the black Gaussian accounts for the third star in the system. The He II 5411 line was fit with a single Gaussian function, and the He II 4686 was fit with a double Gaussian function, but one component in emission. The red line is the resulting best fit model. \label{fig:lines}}
\end{figure*}

\subsection{Radial Velocities} \label{subsec:RV}

In order to derive the radial velocity of the primary, we fit the seven strongest absorption lines with Gaussian functions. To deconvolve the contribution of the primary O-star with that of the third star, we fit a double-Gaussian model, in which one of the Gaussian functions has a fixed velocity (but allowed to be a free parameter) to account for the contribution from the third star, and the other Gaussian has an arbitrary radial velocity, to fit for the velocity of the primary. We do not see evidence for He II lines from the third star. Therefore, we fit the He II 5411 line with a single Gaussian, without including the Gaussian component related to the third star. We also fit the He II 4686 with a double Gaussian, but in this case we constrain one Gaussian function to be positive in order to fit the emission component of this line. An example of the fits to these lines are shown in Figure~\ref{fig:lines}, and the resulting radial velocity curve from fitting each epoch of spectroscopy is shown in the bottom panel of Figure~\ref{fig:model1}.

\subsection{Orbital Period} \label{sec:period}

\begin{figure}
	\centering
	\includegraphics[width=\columnwidth]{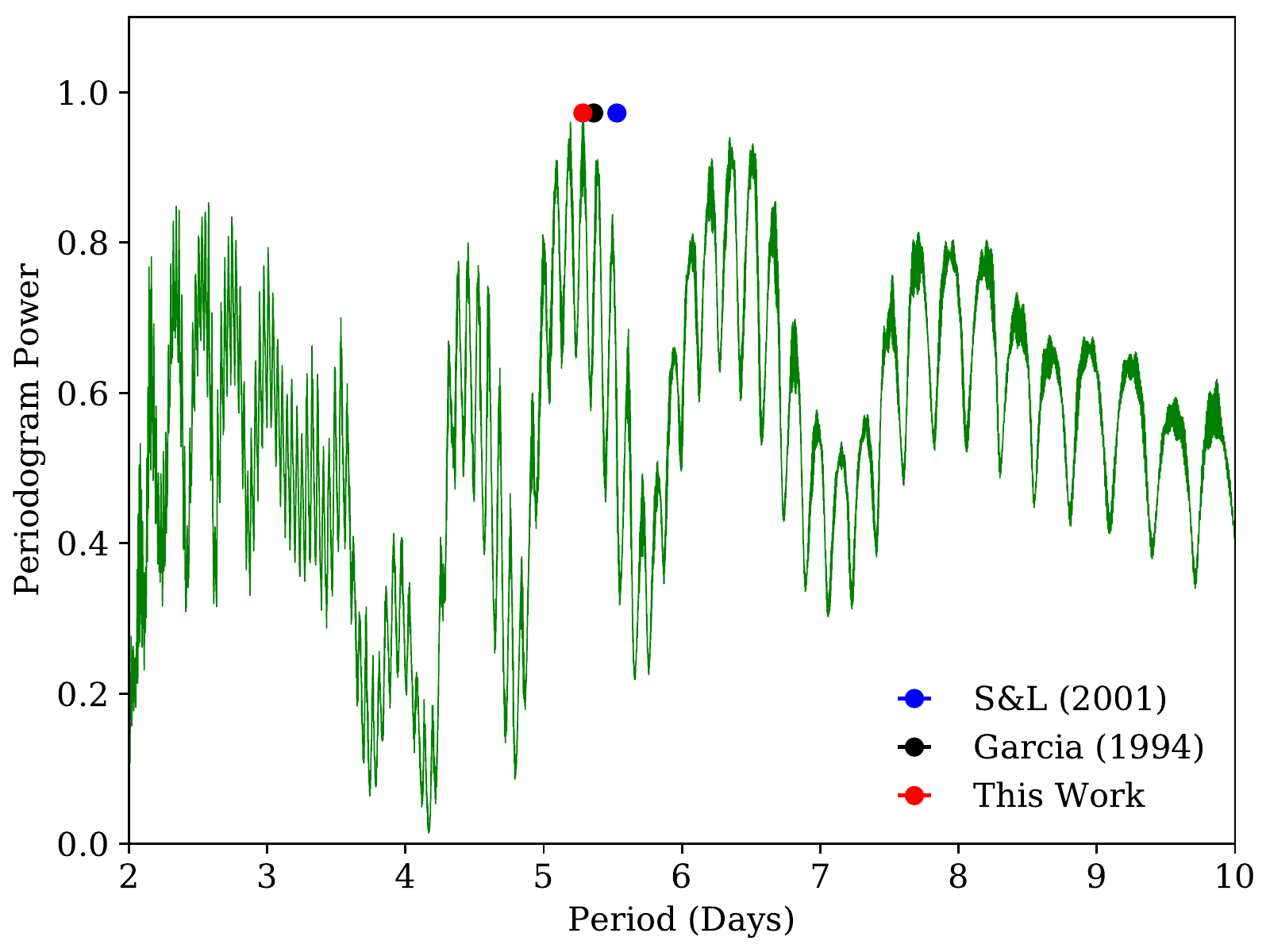}
	\caption{Lomb-Scargle periodogram from all the available radial velocity measurements of HD96670, including measurements from \cite{2001Obs...121....1S} and \cite{Garcia94}. The peak of the periodogram marks the best orbital period of $P = 5.28388 \pm 0.00046$ days. The best periods that other authors determined are shown for comparison, where the measured uncertainties are smaller than the marker size. \label{fig:periodogram}}
\end{figure}

In order to update the orbital period of the binary, we combined our radial velocity measurements with previously published data from \cite{2001Obs...121....1S} and \cite{Garcia94}. These data go back to 1980, effectively increasing our baseline to 37 years. We determine the best orbital period using the {\tt gatspy} python package, an implementation or the Lomb-Scargle periodogram \citep{2015ApJ...812...18V}. The resulting periodogram power spectrum is shown in Figure \ref{fig:periodogram}. We find a best orbital period of $P = 5.28388 \pm 0.00046$ days, a shorter period than the one from \cite{2001Obs...121....1S} of $P = 5.52963 \pm 0.00018$ days, which only uses data up to 1985. The other peaks in the periodogram are aliases, which we can confidently rule out as possible periods through visual inspection of the phased light curve. The light curve phased at this period and a spectroscopic phase 0 of \mbox{$T_0 = {\rm HJD\ } 2444259.549$} is shown in Figure~\ref{fig:photometry}.

\section{Modeling} \label{sec:modeling}

We modeled the optical photometry and radial velocity data simultaneously with the Wilson-Devinney code (hereafter W-D, \citealt{1971ApJ...166..605W, 1979ApJ...234.1054W}), a widely used program that generates synthetic light curve and radial velocity models for a wide variety of close binaries. We model the system as a detached binary where the less massive component is a compact non-radiating object. We assumed the primary to be tidally locked in a synchronous orbit with a fixed orbital period of $5.28388$ days. The remaining parameters we fit for are: The mass of the secondary $M_2$, the mass of the primary $M_1$, the argument of periastron $\omega$, the orbital eccentricity $e$, the surface potential of the primary $P_1$ (A quantity used by the W-D code to calculate the radius of the primary), the orbital inclination $i$, the systemic velocity of the system $\gamma$, and an effective phase shift $\phi$ and $\phi_{RV}$ for the photometry and radial velocity, respectively. We also include a constant third light flux to account for the third star in the system and fix it to $0.31\times$ the flux of the primary O star, based on the flux ratio in H band of $H_{3rd}/H_{O} = 0.31 \pm 0.03$ found by \cite{2014ApJS..215...15S}.

We see that our radial velocity measurements appear to have a constant offset of $16.7 \pm 0.85$ km s$^{-1}$ compared to the measurements from \cite{2001Obs...121....1S}. This is likely due to the presence of the third star in the system, which can affect the apparent systemic velocity of the binary, especially after 37 years. To account for this we also fit for $\Delta \gamma$, a shift in the systemic velocity between our measurements and those of \cite{2001Obs...121....1S}.

\begin{deluxetable}{ccccc}[ht!]
	\tablecaption{Best fit system parameters \label{tab:flaremodels}}
	\tablewidth{0pt}
    \tablehead{Parameter   	  & Model 1                       		& Model 2}
    \startdata
    $M_1\ [M_\odot]$       & $22.0\pm6.5$      & $22.7^{+5.2}_{-3.6}$     \\
    $M_2\ [M_\odot]$       & $4.6 \pm 0.1$     & $6.2^{+0.9}_{-0.7}$      \\
    $\omega_{RV}$ [rad]    & $0.79\pm 0.03$    & $1.10 \pm 0.06$          \\
    $\omega_{Phot}$ [rad]  & \nodata           & $5.91 \pm 0.06$          \\
    $e$                    & $0.28 \pm 0.01$   & $0.12 \pm 0.01$          \\
    $R_1\ [R_\odot]$       & $12.2 \pm 0.2$    & $17.1^{+1.6}_{-1.2}$     \\
    $\phi_{Phot}$          & $0.25 \pm 0.01$   & $0.265 \pm 0.002$        \\
    $\phi_{RV}$            & $0.23 \pm 0.01$   & $0.243 \pm 0.001$        \\
    $\gamma$ [km/s]        & $-26.3 \pm 0.4$   & $-27.50 \pm 0.02$        \\
    $\Delta\gamma$ [km/s]  & $16.58 \pm 0.4$   & $15.87 \pm 0.02$         \\
    $i$ [deg]              & $86 \pm 3$        & $49 \pm 2$               \\
    $F_{H}$ [mag]          & \nodata           & $0.047 \pm 0.002$        \\
    $F_{C}$ [phase]        & \nodata           & $0.798 \pm 0.001$        \\
    $F_{W}$ [phase]        & \nodata           & $0.045 \pm 0.002$        \\
    $\chi^2_{Phot} (dof)$  & $2.20 (43)$       & $0.99 (38) $             \\
    $\chi^2_{RV} (dof)$    & $6.23 (51)$       & $6.06 (49)$              \\
	\enddata
	\tablecomments{The resulting best fit parameters to the radial velocity curve measurements and photometric light curve. Both models were generated with the W-D code assuming an O star primary and non-radiating secondary. Model 1 ignores the excess flux around phase $0.80$, whereas Model 2 models this excess with a Gaussian. Model 2 has two distinct arguments of periastron $\omega$ for the photometry and radial velocity. The parameters listed are: The mass of the primary $M_1$, the mass of the secondary $M_2$, the argument of periastron $\omega$, the eccentricity $e$, the radius of the primary $R_1$, the orbital phase shift $\phi$, the systemic velocity $\gamma$, the difference between our data set and the \cite{2001Obs...121....1S} dataset $\Delta\gamma$, the orbital inclination $i$, and the height, center and width of the Gaussian function that models the excess Flux $F_H$, $F_C$, and $F_W$. The reduced $\chi^2$ values for each model are shown, separating the value for the radial velocity and photometry. The plots corresponding to the best fit models to the data are shown in Figures \ref{fig:model1} and \ref{fig:model4}.}
\end{deluxetable}

We run the W-D code using a custom Python wrapper to be able to fit the data with a Markov chain Monte Carlo (MCMC) approach using the {\tt emcee} \citep{2013PASP..125..306F} implementation of the Goodman and Weave \citep{Goodman10} algorithm. We fit the inclination with a prior that is flat in $\cos(i)$, and the $\Delta\gamma$ with a Gaussian prior of $16.7 \pm 0.85$ km s$^{-1}$ (the difference between the average radial velocity measurements from our work and those from \cite{2001Obs...121....1S}). We adopt flat uninformative priors for all other parameters. We run the sampler with 300 walkers each taking 1000 steps and discarding the first 40\% for burn-in. The W-D code implementation is computationally expensive and hard to parallelize, which results on computation times in the order of $\sim10$ days on a typical personal computer. We test for convergence by using the Gelman-Rubin statistic and see that the potential scale reduction factor is $\hat{R} < 1.5$ \citep{1992StaSc...7..457G}.

\begin{figure}
    \includegraphics[width=\columnwidth]{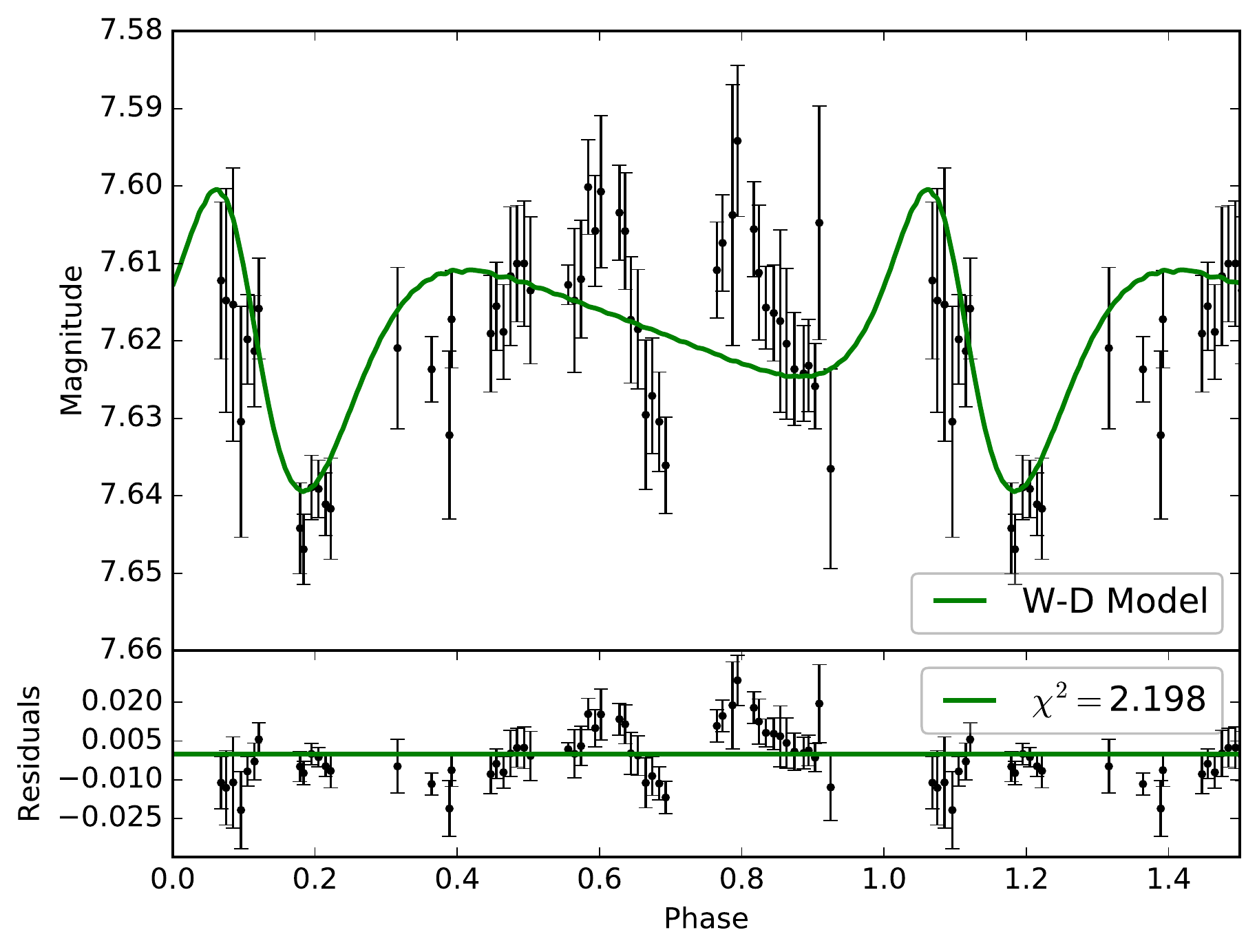}
    \includegraphics[width=\columnwidth]{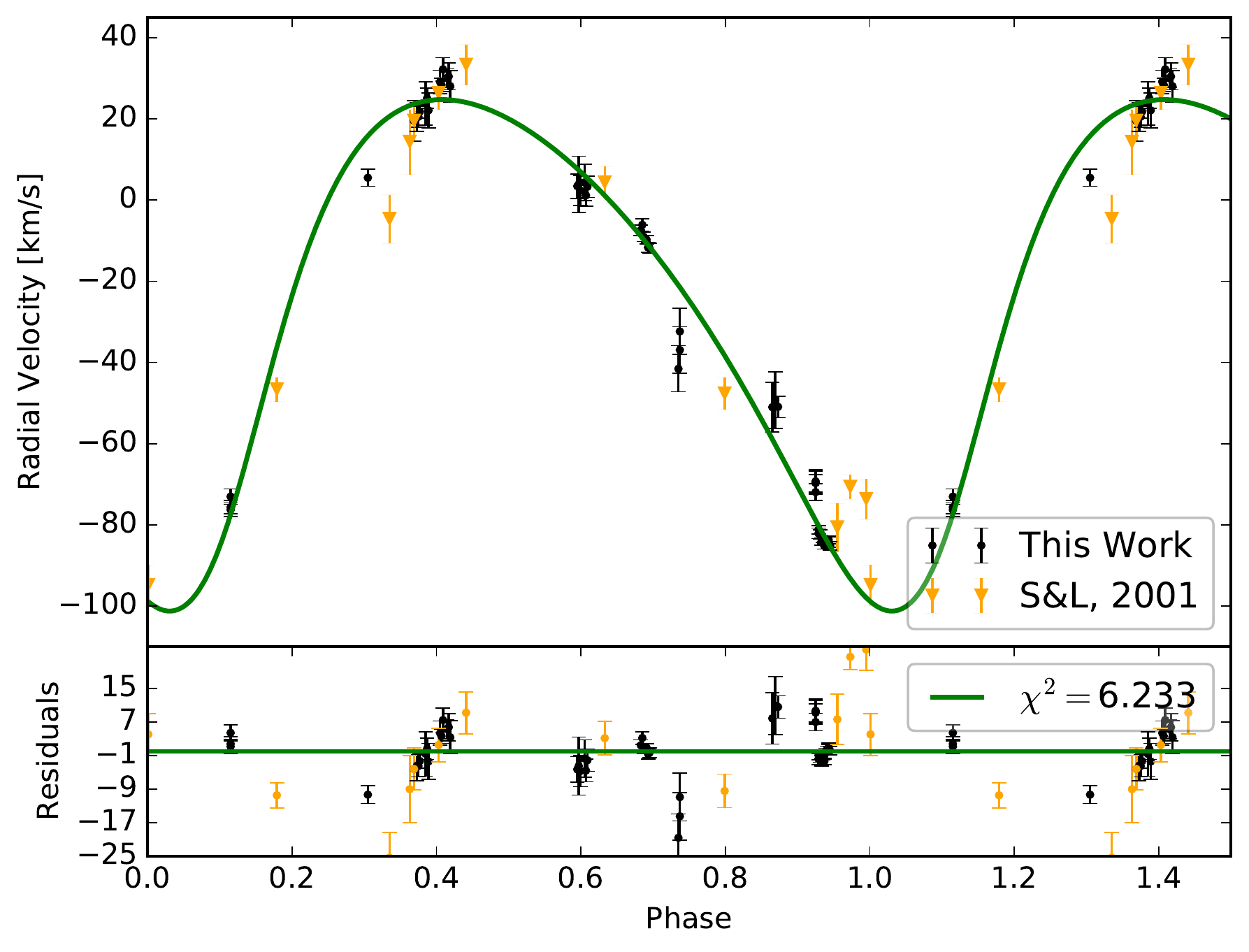}
	\caption{Model 1: Simplest W-D model fit for the phased light curve (top) and radial velocity curve (bottom). The parameters of this model are shown in Table \ref{tab:flaremodels}. This model makes no extra assumptions about about the excess flux around phase $0.75 - 0.95$, and fails to accurately represent the data around that region. The orange points in the radial velocity curve are measurements from \cite{2001Obs...121....1S}. \label{fig:model1}}
\end{figure}

\begin{figure*}
    \centering
	\includegraphics[width=\textwidth]{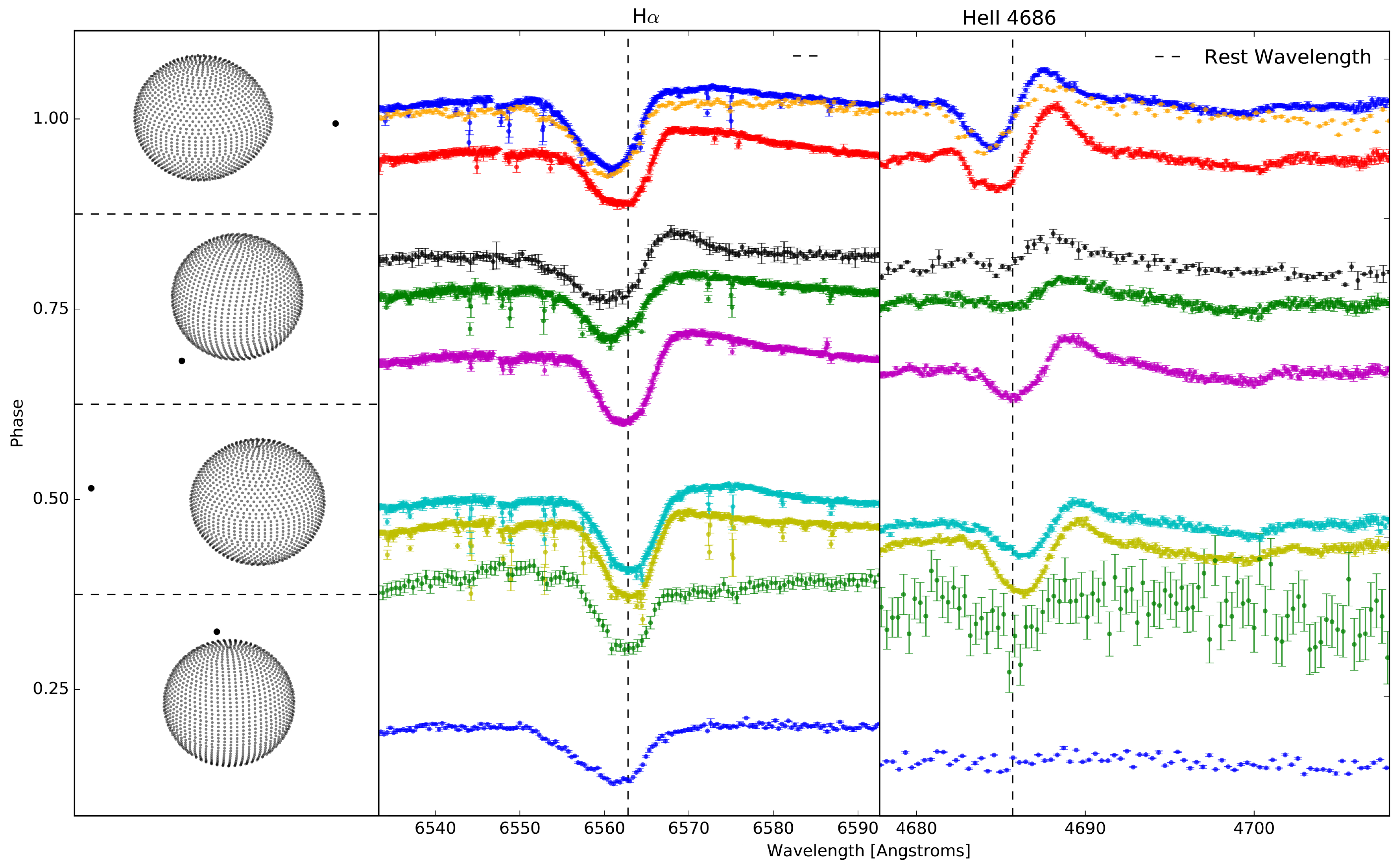}
	\caption {The left panels show a schematic representation of the binary generated by the W-D code using the best fit parameters from Model 2, shown in Table~\ref{tab:flaremodels}. The center and right panels show H$\alpha$ and He II 4686, two strong emission lines with orbital dependence, plotted as a function of orbital phase in the y-axis. Each color represents a different spectrum. \label{fig:halpha}}
\end{figure*}

First, we fit the light curve and radial velocity data simultaneously with the model described above with no additional assumptions. The best fitting model is shown in Figure~\ref{fig:model1} with the list of parameters for Model 1 listed in Table~\ref{tab:flaremodels}. We see that this simple model fails to accurately represent the data, especially the light curve around phase $\sim 0.8$, this is due to an excess flux that deviates from a normal sine curve (See Figure~\ref{fig:photometry}). This excess flux is present in all bands, observed on different epochs, and is therefore not a one time flare-like event. This is the region in phase where the secondary accretor is in inferior conjunction and transiting the O star (see Figure \ref{fig:halpha} for a schematic representation of the binary). We interpret this excess flux as either coming from a very narrow focused wind from the primary accreting onto the secondary, a small hot spot located at the pole of the primary, or as a reduced opacity in the accretion wind due to the focused wind shocks and the accretion bow shock that causes an apparent brightening of the primary. In order to account for this excess flux, in the following model (Model 2) we add a simple Gaussian function of height $F_H$, center $F_C$, and width $F_W$ to the light curve that is meant to represent a general model for any of the possibilities outlined here.

\begin{figure}
    \includegraphics[width=\columnwidth]{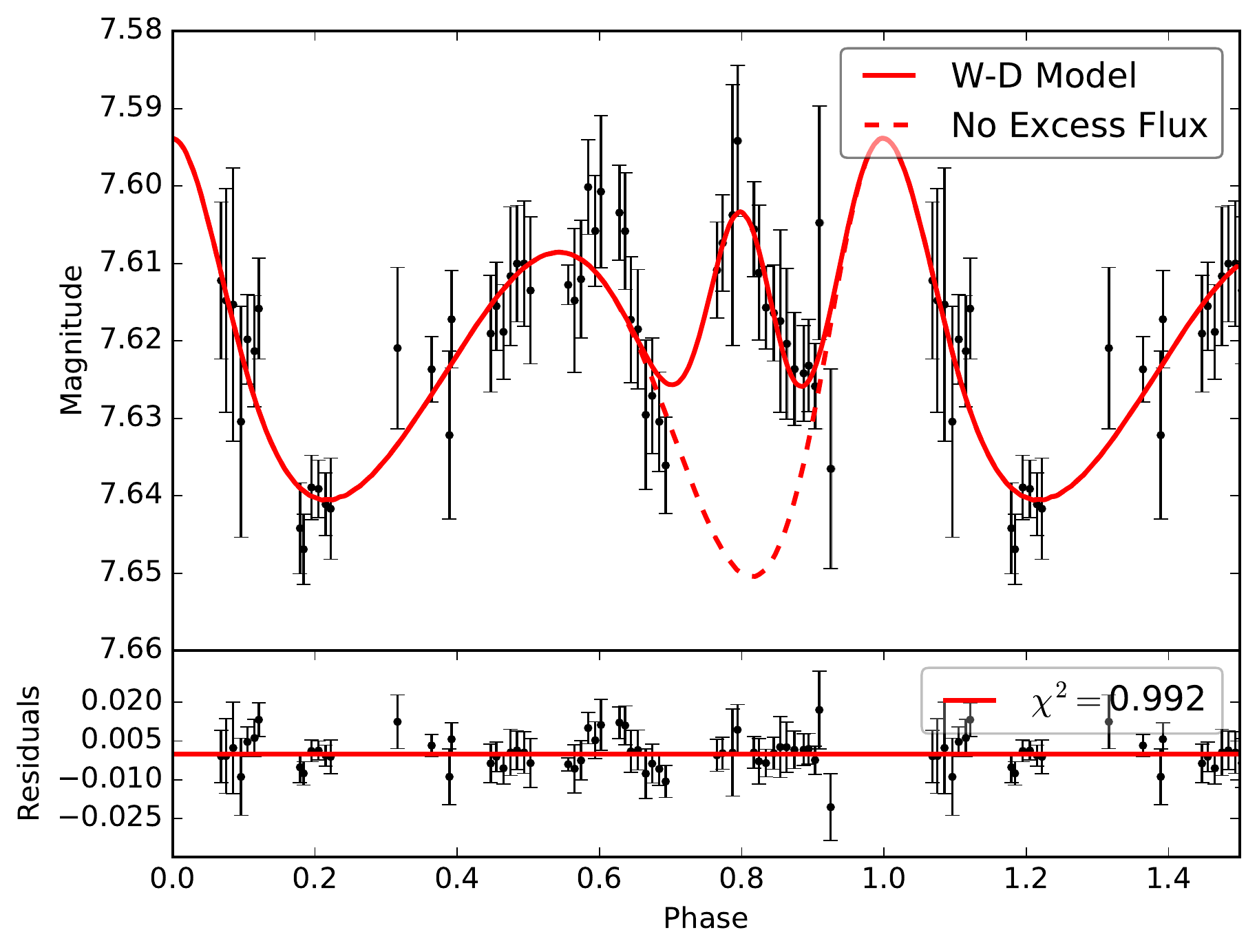}
    \includegraphics[width=\columnwidth]{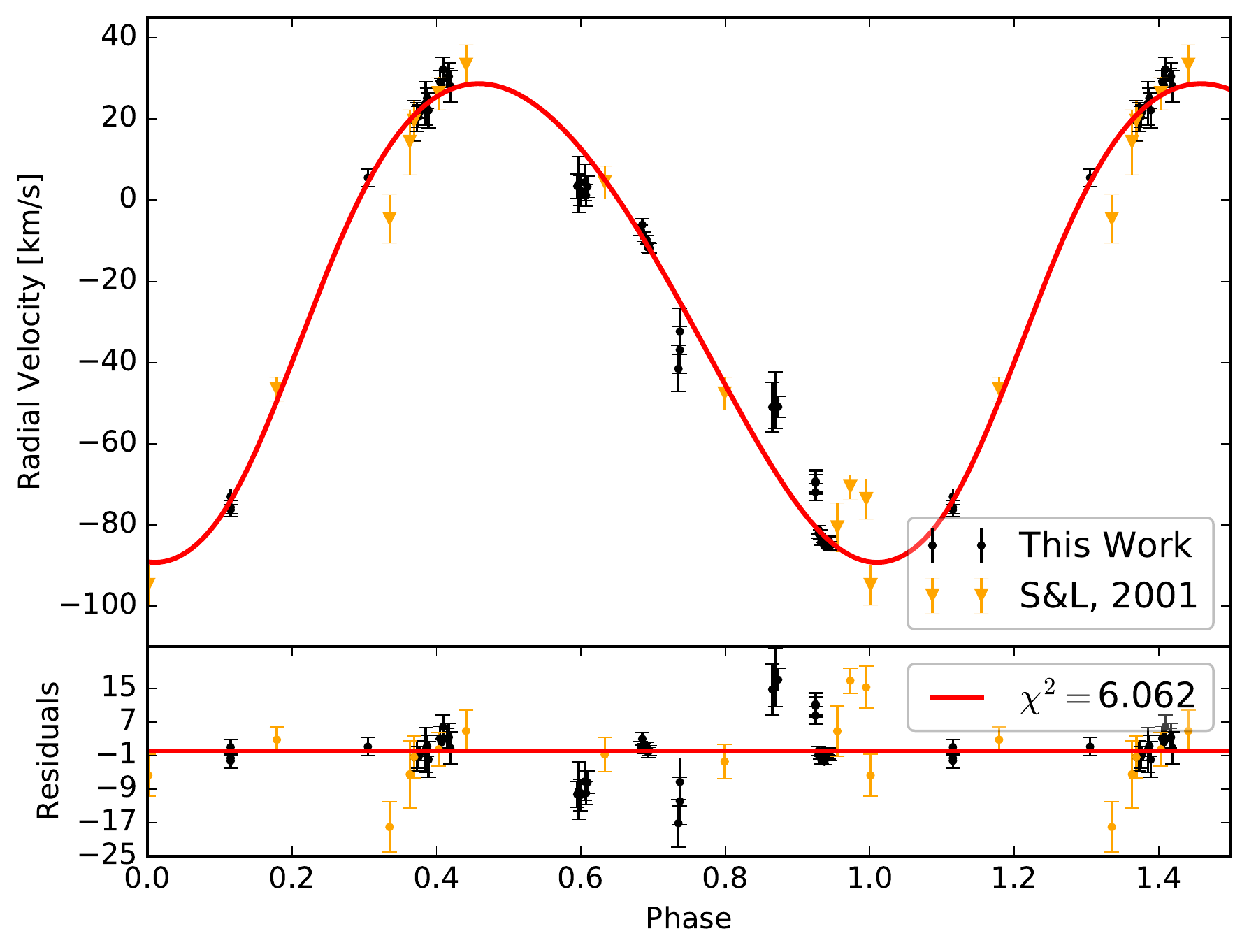}
	\caption{Model 2: Best W-D model fit for the light curve (top) and radial velocity curve (bottom), the parameters for this model are shown in Table \ref{tab:flaremodels}. This model fits the region in phase $0.75 - 0.95$ with a simple Gaussian and fits two independent arguments of periastron for the radial velocity and photometric data. The orange points in the radial velocity curve are measurements from \cite{2001Obs...121....1S}. \label{fig:model4}}
\end{figure}

\begin{figure*}
    \centering
	\includegraphics[width=0.8\textwidth]{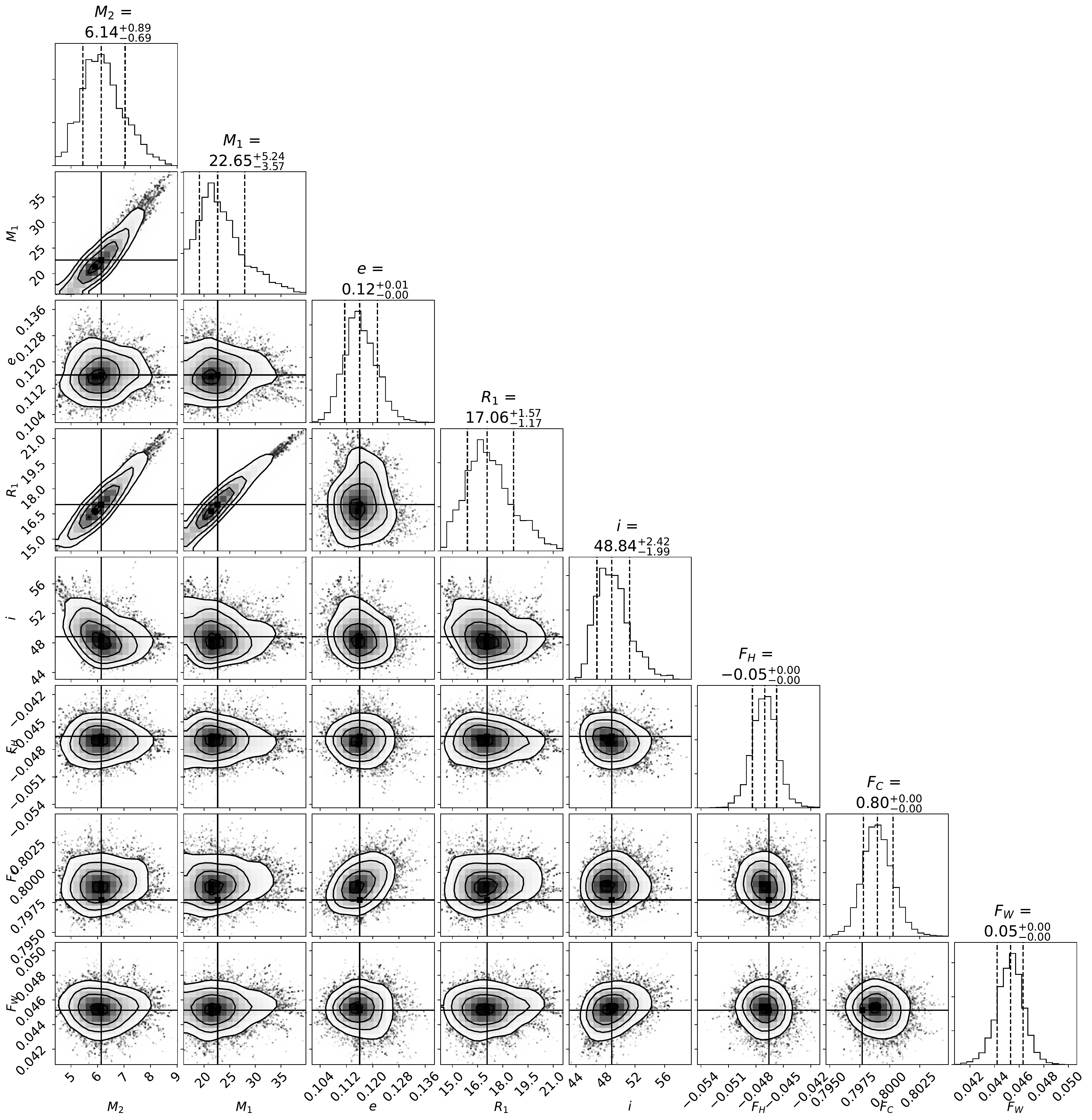}
	\caption{Sample results of a Markov chain Monte Carlo run for the possible models for the binary in HD96670. We show the two-dimensional correlation matrices for the most important physical parameters. The diagonal shows the marginalized posterior distribution of each parameter, with vertical lines at the best parameter estimate and $1\sigma$ levels. Most parameters are well behaved and uncorrelated, with the exception of the primary and secondary mass and the primary radius. This figure was generated with the {\tt corner} Python package \citep{Foreman16}. \label{fig:corner}}
\end{figure*}

Aditionally, we test a model in which we fit the light curve and radial velocity curve separately. When doing this we notice that the best-fit argument of periastron is different for the two data sets. Therefore, in Model 2 we fit for two independent arguments of periastron, one for the photometry $\omega_{Phot}$, and one for the radial velocity $\omega_{RV}$. The tidal distortion of stars in X-ray binaries can have an effect on the measured eccentricity or argument of periastron of the radial velocity curve, especially for high mass systems. \cite{1976ApJ...203..182W} determined that the larger the mass ratio of a binary, the less sinusoidal the radial velocity curve will be, since the tidal distortion of the primary produces a net asymmetry in the disk of the star this will have an effect in the rotational velocity and could be the cause behind spurious measurements of parameters such as the argument of periastron \citep{2008ApJ...681..562E}. Nevertheless, this should not be of great concern, since the rest of the more relevant physical parameters such as the mass of the secondary $M_2$, primary $M_1$, eccentricity $e$, surface potential of the primary $P_1$, and orbital inclination $i$ are all in agreement between the radial velocity curve and light curve and remain consistent when fitted together under the assumptions of Model 2.

The list of best fit parameters from Model 2 are shown in Table~\ref{tab:flaremodels}, with the best fit model shown Figure~\ref{fig:model4}, and the corresponding two-dimensional correlation plot for the most relevant parameters in Figure~\ref{fig:corner}. From our fits we see that most parameters are Gaussian and uncorrelated, except for the primary mass $M_1$, which is degenerate with the secondary mass $M_2$ and primary radius $R_1$. We measure a strong correlation of $M_2 = 0.185 M_1 + 1.958$ and $R_1 = -0.00478 M_1^2 + 0.5433 M_1 + 7.3805$. The posterior probability distribution for the primary is centered at $M_1 = 22.7^{+5.2}_{-3.6} M_\odot$, which we adopt as the most likely mass estimate.

\section{Discussion} \label{sec:Discussion}

From Model 2 we determined the mass of the accretor to be $M_2 = 6.2^{+0.9}_{-0.7}$ M$_\odot$. This result rules out a white dwarf or neutron star companion, which leaves as possible accretors are either a B-type star or a black hole. The fact that we see no evidence of absorption lines from the companion helps to rule out bright stars $\gtrsim 10 M_\odot$, nevertheless we can not rule out a B-type star secondary with a mass lower than this based solely on the lack of absorption lines from a secondary star. If the companion is not a black hole it could be a B3-B7 type main sequence star. We find the best radius for the primary to be $17.1^{+1.6}_{-1.2}$ R$_\odot$, which is consistent with either a late-stage main sequence O8V star, or a sub-giant O8IV that is just turning off the main sequence. Additionally, we see that the orbit of HD96670 shows some orbital eccentricity, which could be induced by the presence of the third star companion. For the parameters of this system, Kozai-Lidov oscillations are expected on a timescale of $1-2\times10^{6}$ yr \citep{Toonen20}, within the life span of HD96670.

We measure the equivalent width of He I 4471 and He I 4542 in our 10 individual IMACS spectra and find an average ratio of $\log(EW(4471)/EW(4542)) = 0.227 \pm 0.007$, using the correlation from \cite{2018arXiv180508267M}, we determine this is most consistent with an O8.5 star, in agreement with the spectral classification from \cite{2014ApJS..211...10S}.

The measured separation between the main binary and the third star companion is $\rho = 29.9 \pm 3.37$ mas \citep{2014ApJS..215...15S}, or a physical projected separation of $47 \pm 6$ AU (for a distance of 3.87 kpc). In order to estimate the actual physical separation between the binary and the third star we perform a statistical analysis using a Monte Carlo approach by randomly sampling from a distribution of orbital inclinations uniform in $\cos(\theta)$, where $\theta$ is the angle between the plane of the main binary and the third star. This, in combination with the projected separation and distance to the HD96670 yields a physical separation of $a = 56^{+31}_{-10}$ AU. We can estimate the orbital period of the third star using simple Newtonian mechanics. Assuming the total mass of the three stars is $M_T = 28.29 \pm 2.03$, that would imply an orbital period of $P = 91^{+29}_{-23}$ yr. We can not confirm the orbit of the third star since we only have two measurements for the systemic velocity of the main binary, one from this work ($\gamma = -27.33  \pm 0.44$  km/s) and one from \cite{2001Obs...121....1S} ($\gamma = -9.0 \pm 1.7$ km/s), separated by 37 years, or separated in phase by $0.41^{+0.14}_{-0.10}$.

The measured velocities of the main binary ($\gamma = -27.33  \pm 0.44$ km/s) and the third star ($v = 39.3 \pm 0.3$ km/s) are such that their relative velocity is $v_r = 67.23 \pm 0.53$ km/s. If gravitationally bound, we can derive the orbital velocity for the third star using the orbital period $P$ and separation $a$ of the third star calculated above. The orbital velocity of the third star would be $27^{+20}_{-15}$ km/s, or $2\sigma$ lower than the measured relative velocity. This suggests that the third star might not be gravitationally bound to the main binary.

In Figure~\ref{fig:halpha} we show a schematic representation of the best-fit W-D model compared to the H$\alpha$ and He II 4686 lines at the corresponding phase. We interpret the emission component in the spectra as being caused by the focused wind of the primary. The emission component of both H$\alpha$ and He II are strongest when the secondary is close to inferior conjunction, which happens at phase = 0.793 due to the eccentricity of the system. The lack of He II absorption at phase $\sim 0.75$ can be explained by strong He II emission at this phase that is ``filling in" the He II absorption.

We note a peculiar feature from the He II 4686 lines, which show an emission component redshifted by a constant $\sim 200$ km s$^{-1}$, seen at both phases $\sim 0.9$ and $\sim 0.5$. One possibility is that this is caused by a complex accretion wind that is curved by the orbit of the binary where the blueshifted emission is eclipsed or otherwise absorbed around inferior conjunction by a denser wind. But further observations are needed to track the evolution of this line. The presence of redshifted emission and lack of blueshifted emission are not consistent with a simple accretion wind model and this peculiar He II feature remains mostly unexplained.

\section{Conclusion} \label{sec:conclusions}

We have presented analysis and modeling of a new BH-HMXB candidate, the single-line spectroscopic binary HD96670. The primary donor star is an O8 main sequence or sub-giant type star with a mass of $22.7^{+5.2}_{-3.6} M_\odot$ being orbited by a secondary of $6.2^{+0.9}_{-0.7} M_\odot$. We can rule out a white dwarf or neutron star secondary based on the minimum mass of the secondary, leaving a black hole or a B3-B7 main sequence star as possible companions. The hard X-ray spectra are not easily explained with a B-type companion, and a black hole accretor is therefore favored. More optical photometry is needed to average out the large optical variations and flaring that exists in the system. More optical spectroscopy is needed to better sample the radial velocity curve, and to understand the nature behind the He II 4686 emission component. Coordinated Chandra or XMM-\textit{Newton} and NuSTAR observations would aid towards mapping out the thermal Fe line emission versus power law continuum to better constrain the wind and accretion geometries in order to determine if, in fact, HD96670 harbors a compact accretor.

It is natural to compare HD96670 to Cygnus X-1, the only undisputed BH-HMXB in the galaxy. Cygnus X-1 has an O9.7Iab supergiant primary with a mass of $M_1 = 19.2 \pm 1.9 M_\odot$ and a radius of $R_1 = 16.17 \pm  0.68 R_\odot$ and a black hole accretor with a mass of $M_2 = 14.8 \pm 1.0 M_\odot$ on a $5.599829$ day orbital period and an orbital inclination of $i = 27.1 \pm 0.8^\circ$ \citep{2011ApJ...742...84O,2016A&A...590A.114M}. This equates to a mass function of $f(M) = 0.265 \pm 0.053 M_\odot$. The orbital period and primary mass of Cygnus X-1 are similar to HD96670. If the secondary in HD96670 is a black hole, this would make it a Cygnus X-1 progenitor system, allowing us to study the evolution of HMXBs by linking Cygnus X-1 to its early life stages.

\section{Acknowledgments} \label{sec:acknowledgments}

We thank the anonymous referee for useful comments towards the improvement of this paper. This research is supported in part by an NSF Graduate Research Fellowship and NSF grant AST1313370. This work makes use of data from the American Association of Variable Star Observers, we acknowledge the variable star observations from the AAVSO International Database contributed by observers worldwide and used in this research. This paper includes data gathered with the 6.5 meter Magellan Telescopes located at Las Campanas Observatory, Chile. This research has made use of NASA’s Astrophysics Data System. This research has made use of the SIMBAD database, operated at CDS, Strasbourg, France. This research has made use of NASA’s Astrophysics Data System.

\software{PyRAF\citep{science12}, SAOImage DS9 \citep{Smithsonian00}, Astropy \citep{astropy}, Matplotlib \citep{matplotlib}, emcee\citep{2013PASP..125..306F}, corner \citep{Foreman16}, NumPy \citep{numpy}, Wilson-Devinney \citep{1971ApJ...166..605W}}

\bibliography{references}

\end{document}